\documentclass{iaus}
\usepackage{graphicx}

\title[The rapid magnetic rotator HR\,7355] 
      {The rapid magnetic rotator HR\,7355\footnote{Based on observations
          under ESO programs 081.D-2005, 383.D-0095.}}

\author[Th.\ Rivinius et al.]   
       {Th.\ Rivinius$^{1}$, R.H.D.~Townsend$^2$, S.~\v{S}tefl$^{1}$,
         D.~Baade$^3$
}

\affiliation{$^1$ESO, Chile; 
$^2$UW Madison, USA;
$^3$ESO, Germany
}

\pubyear{2011}
\volume{272}  
\pagerange{1--6}
\jname{Active OB stars: structure, evolution, mass loss and critical limits}
\editors{C.\ Neiner, G.\ Wade, G.\ Meynet \& G.\ Peters, eds.}
\begin{document}
\maketitle

\begin{abstract}
For early type magnetic stars slow, at most moderate rotational velocities
have been considered an observational fact. The detection of a multi-kilogauss
magnetic field in the B2Vpn star with $P\approx0.52\,d$ and $v\sin i\approx
300$\,km/s has brought down this narrative. We have obtained more than 100
high-resolution, high-S/N echelle spectra in 2009. These spectra provide the
most detailed description of the variability of any He-strong star to
date. The circumstellar environment is dominated by a rotationally locked
magnetosphere out to several stellar radii, causing hydrogen emission. The
photosphere is characterized by surface chemical abundance inhomogeneities,
with much stronger amplitudes, at least for helium, than slower rotating stars
like $\sigma$\,Ori\,E.  The highly complex rotational line profile modulations
of metal lines are probably a consequence the equatorial gravity darkening of
HR\,7355, and thus may offer an independent measurement of the von Zeipel
parameter $\beta$.
\end{abstract}

\firstsection 
\section{Introduction}

The Helium-strong star HR\,7355 has recently been found to host a strong
magnetic field. The short period of $P=0.52$\,d puts it among the shortest
period non-degenerate magnetic stars known, while its $v\sin i \approx
310$km\,s\,$^{-1}$ is the highest for any known non-degenerate magnetic star.

HR\,7355 is a star for which effects like gravity darkening and oblate
deformation cannot be ignored anymore. This means traditional analysis methods
to derive stellar parameters will, at best, give uncertain results, and at
worst misleading ones.

The presence of photospheric abundance pattern is intriguing, as in a hot star
rotationally induced meridional flows should dilute such pattern quickly.
While it has been suggested that the magnetic field would inhibit this
circulation, HR\,7355 is the first case where this can be tested
observationally: not only are there abundance variations across the stellar
surface, but the amplitude of the equivalent width of the He{\sc i} lines is
much larger that in the similar, though less rapidly rotating, Bp star
$\sigma$\,Ori\,E.
\section{Observations}
Apart from archival and literature data this work is based on high-resolution
echelle spectra obtained in 2009 with UVES at the 8.2m Kueyen telescope on
Cerro Paranal. The instrument was used in its DIC2 437/760 setting, which
gives a blue spectrum from about 375 to 498\,nm and a red spectrum from about
570 to 950\,nm, with a small gap at 760\,nm. The slit-width was 0.8\,arcsec,
giving a resolving power of about $R=50\,000$ over the entire spectrum. The
UVES observations were done on three chunks, in April, July, and September
2009.

Exposure times were between 120 seconds in April and 30 seconds July to
September, where the shorter exposures were repeated four times. Since the
period is short we did not average shorter exposures, but treat them as taken
at distinct phases.  In total, we have 104 blue and 112 red spectra. The
typical $S/N$ of a single spectrum is about 275 in the blue region and 290 in
the red.

\begin{figure}
\includegraphics[angle=270,width=\textwidth,clip=]{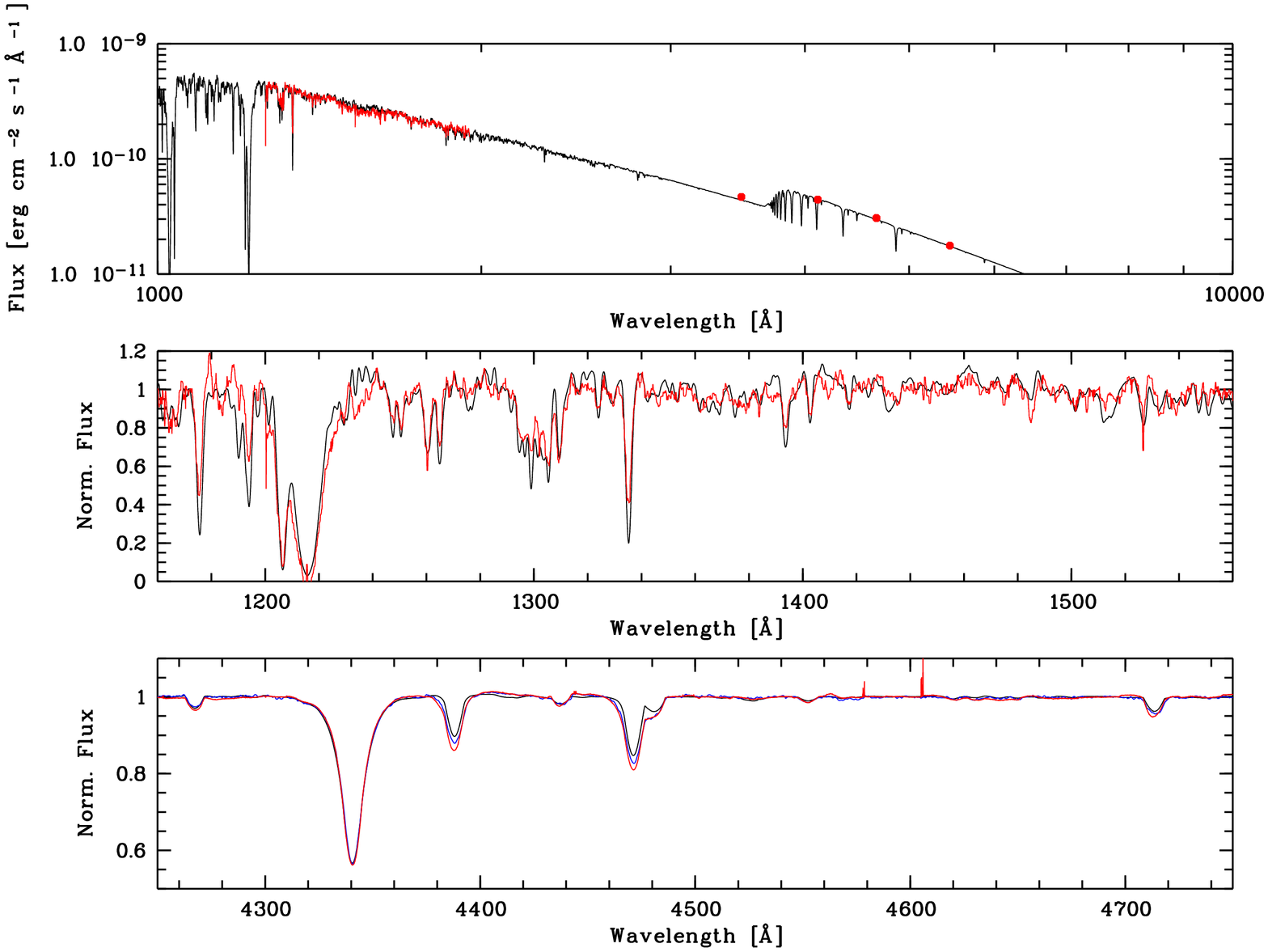}
\caption[]{
\centering
The B3-model vs.\ observed data: Upper panel: UV IUE
  short-wavelength spectrophotometry and Str\"omgren visual photometry
  vs.\ modeled fluxes. Middle panel: IUE spectrum vs.\ modeled UV-line
  profiles. Lower panel: All UVES (red) and FORS2 (blue) spectra averaged
  vs. modeled visual line profiles.}
\label{fig_fluxfit}
\end{figure}

\section{Stellar parameters}

In order to determine the stellar parameters, we made use of spectral
synthesis, both for line strength and profiles, as well as for absolute
fluxes. The code we used is the third, re-programmed version of Townsend's
(\cite[1997]{1997MNRAS.284..839T}) BRUCE and KYLIE suite, which in the further
we will call ``B3''.

In a first step, the profile of {C}{\sc ii}\,4267, the strongest of all metal
lines, was used to obtain the {\bf projected rotational velocity} of $v \sin i
= 310\pm5$\,km\,s$^{-1}$, in good agreement with \cite[Oksala et
  al.\ (2010)]{2010MNRAS.405L..51O}.  Together with the rotational period of
$P=0.5214404$\,d the {\bf equatorial radius} then becomes $R_{\star, \rm equ.}
\sin i = 3.19 {\rm R_{\odot}}$. Assuming that a rapidly rotating star is
constrained by five independent parameters, usually $v_{\rm eq}, i, T_{\rm
  eff}, M_\star, R_\star$, this leaves only three of them to be determined,
namely the mass, the effective temperature, and the inclination.

\begin{figure}
\begin{center}
\includegraphics[viewport=0 105 546 746,angle=0,width=9cm,clip=]{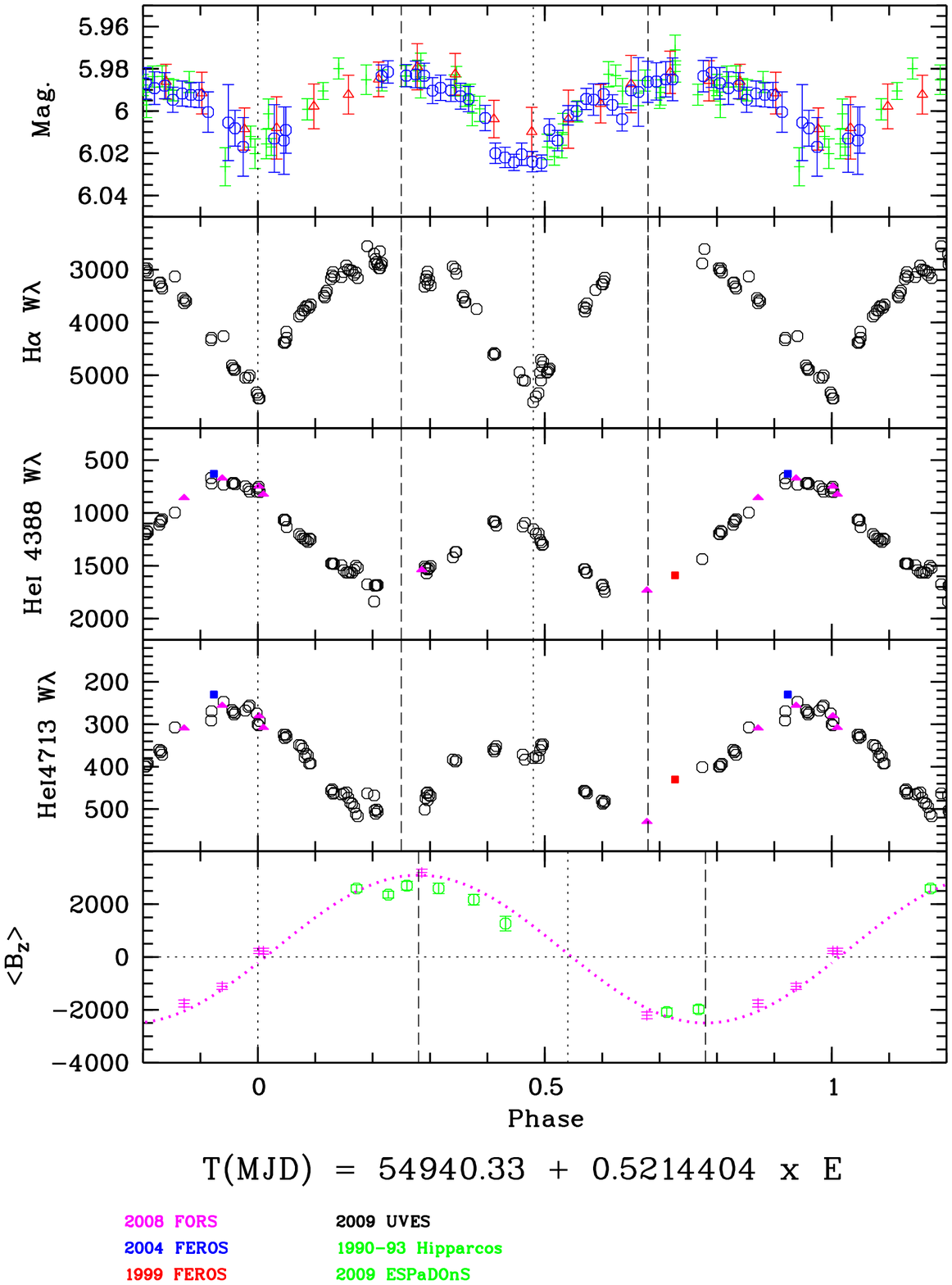}
\end{center}

\caption[]{\centering Phased observational data: Photometric data from
  \cite[Oksala et al.(2010)]{2010MNRAS.405L..51O}, and ASAS data from
  \cite[Mikul{\'a}{\v s}ek et al.(2010)]{2010A&A...511L...7M} (uppermost), EWs
  from UVES (middle), with FORS and FEROS $W_\lambda$ measurements of He{\sc
    i}\,4388, 4713 added, as well as FORS \cite[(Rivinius et
    al.\ (2010))]{2010MNRAS.405L..46R} and ESPaDOns magnetic data (lowermost
  panel).}
\label{fig_phasingall}
\end{figure}

\begin{figure}
\begin{center}
\parbox{0.02\textwidth}{~}%
\parbox{0.25\textwidth}{\centerline{[He{\sc i}]\,4045}}%
\parbox{0.25\textwidth}{\centerline{He{\sc i}\,4713}}%
\parbox{0.25\textwidth}{\centerline{He{\sc i}\,6678}}%
\parbox{0.25\textwidth}{\centerline{Ne{\sc i}\,6717}}%

\includegraphics[viewport=100 188 608 1010,angle=0,width=0.25\textwidth,clip=]{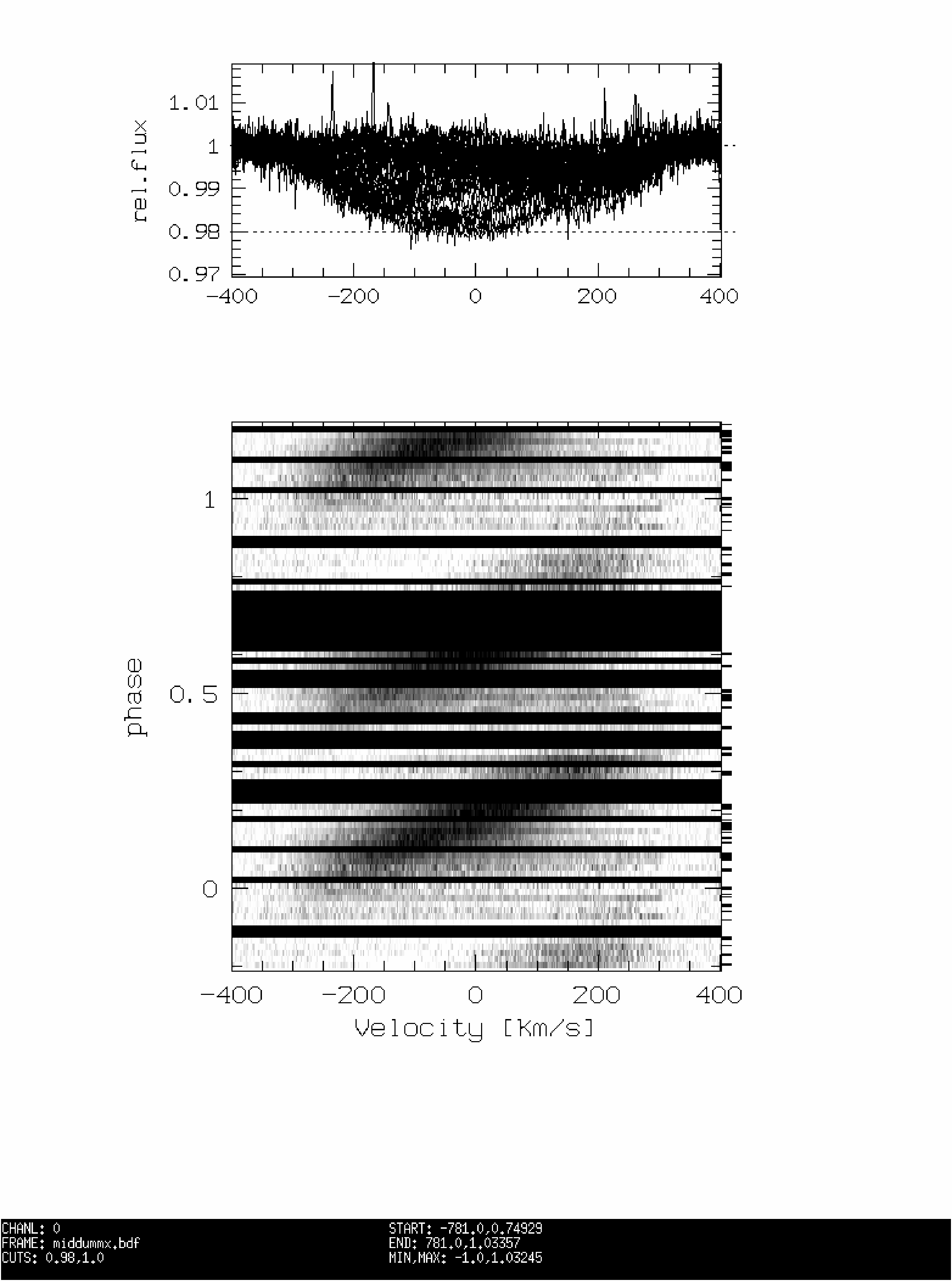}%
\includegraphics[viewport=100 188 608 1010,angle=0,width=0.25\textwidth,clip=]{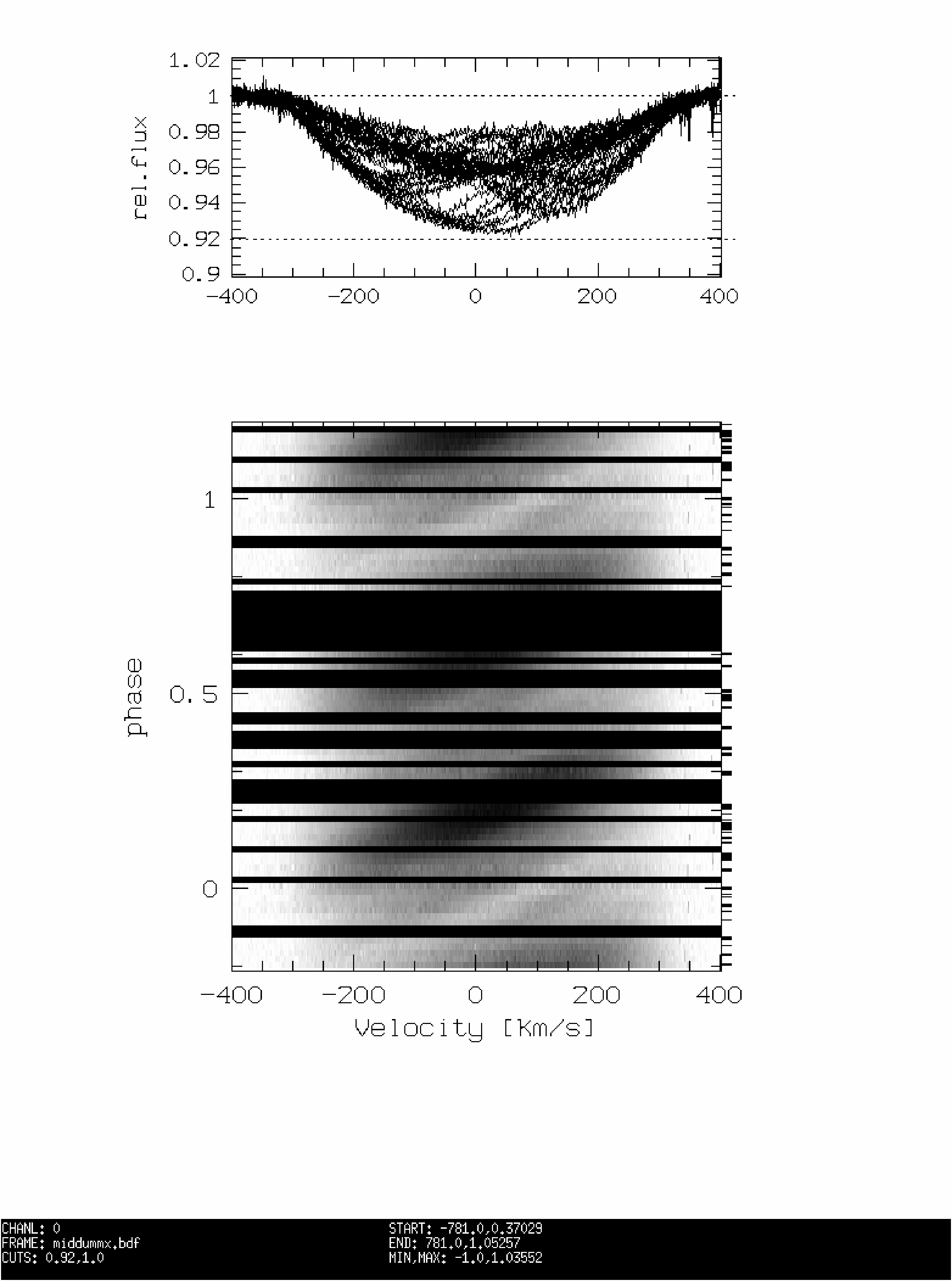}%
\includegraphics[viewport=100 188 608 1010,angle=0,width=0.25\textwidth,clip=]{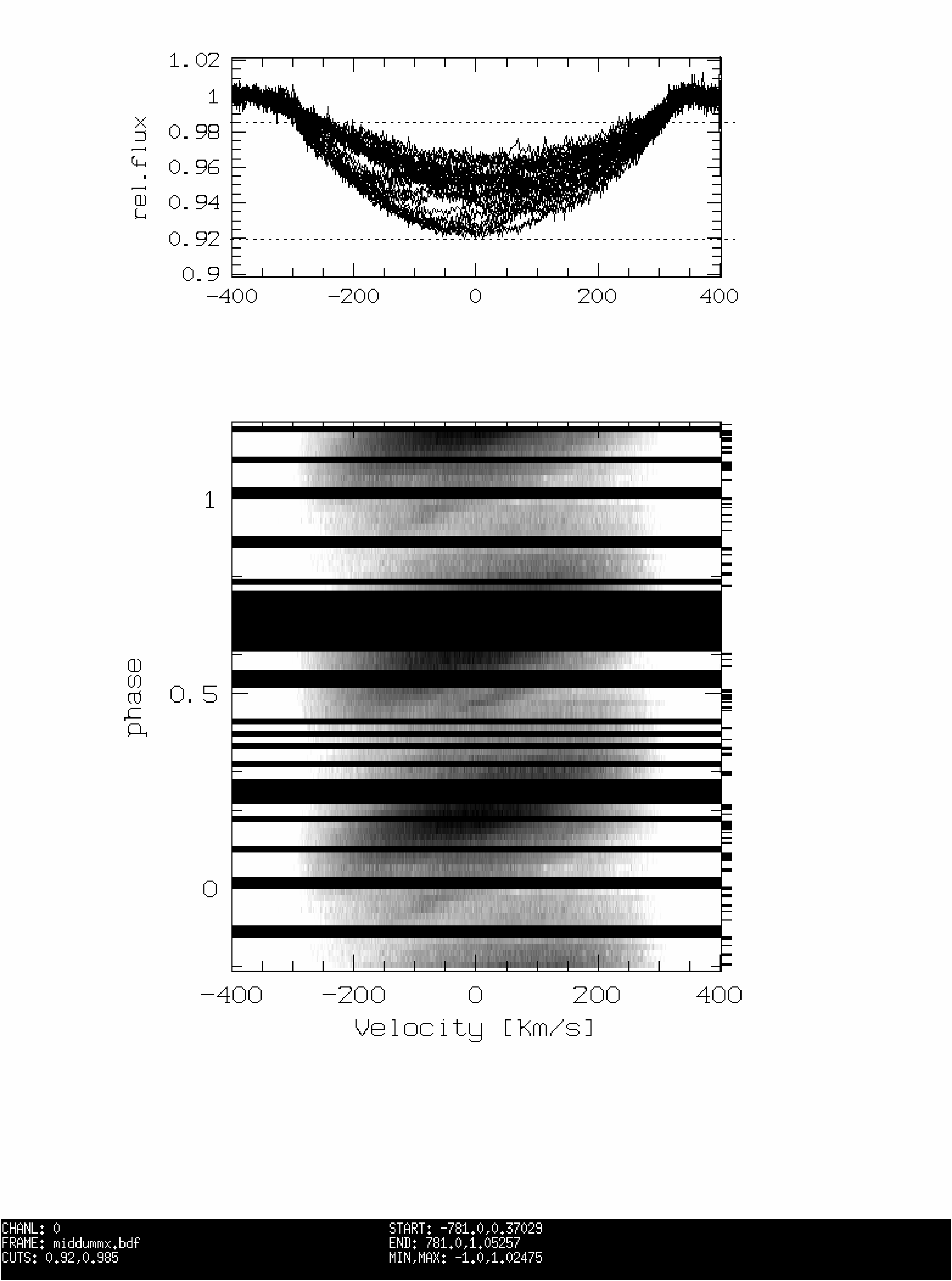}%
\includegraphics[viewport=100 188 608 1010,angle=0,width=0.25\textwidth,clip=]{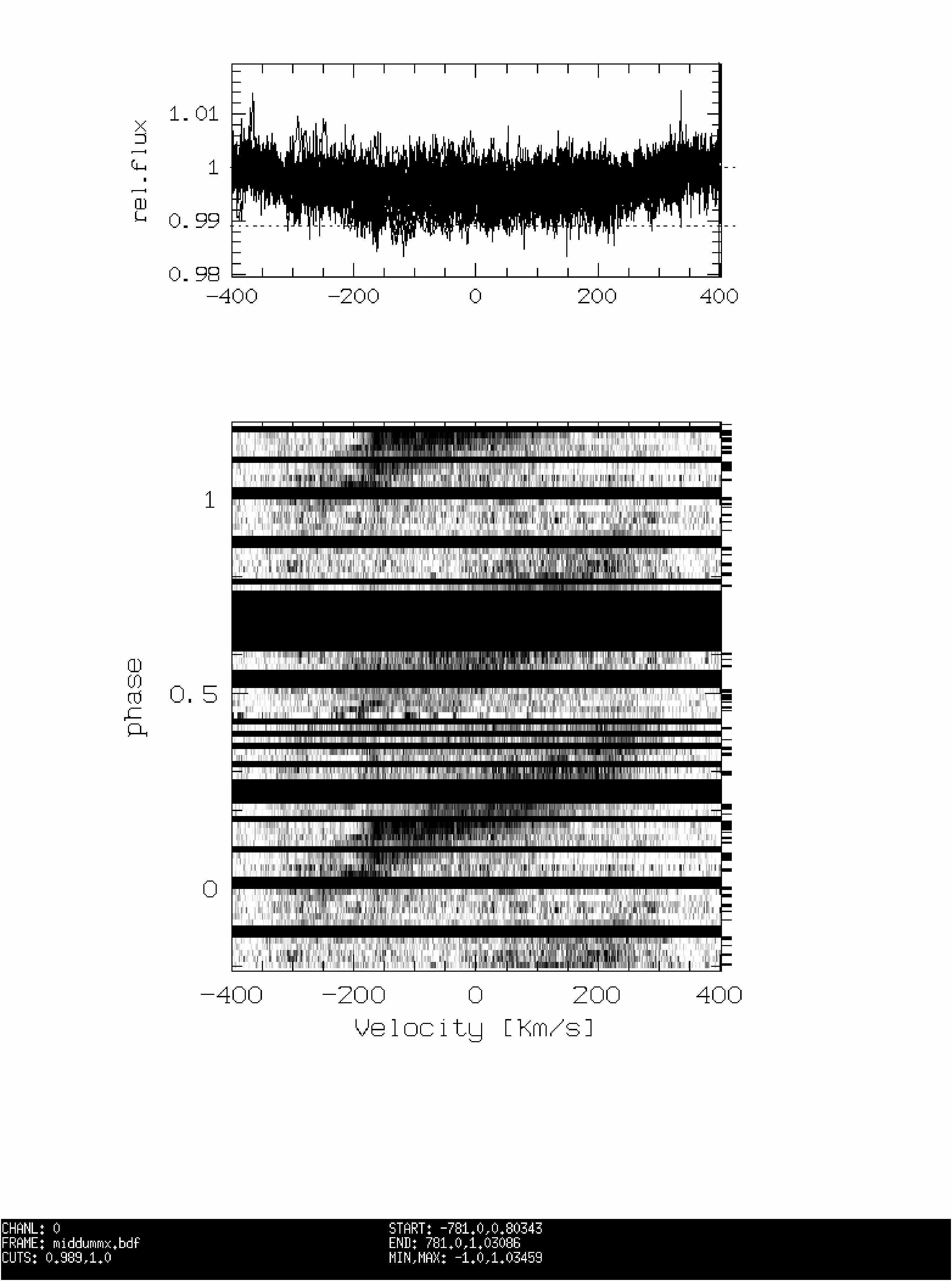}%

\parbox{0.02\textwidth}{~}%
\parbox{0.25\textwidth}{\centerline{N{\sc ii}\,4631}}%
\parbox{0.25\textwidth}{\centerline{C{\sc ii}\,4267}}%
\parbox{0.25\textwidth}{\centerline{Si{\sc iii}\,4553}}%
\parbox{0.25\textwidth}{\centerline{S{\sc ii}\,4816}}%

\includegraphics[viewport=100 188 608 1010,angle=0,width=0.25\textwidth,clip=]{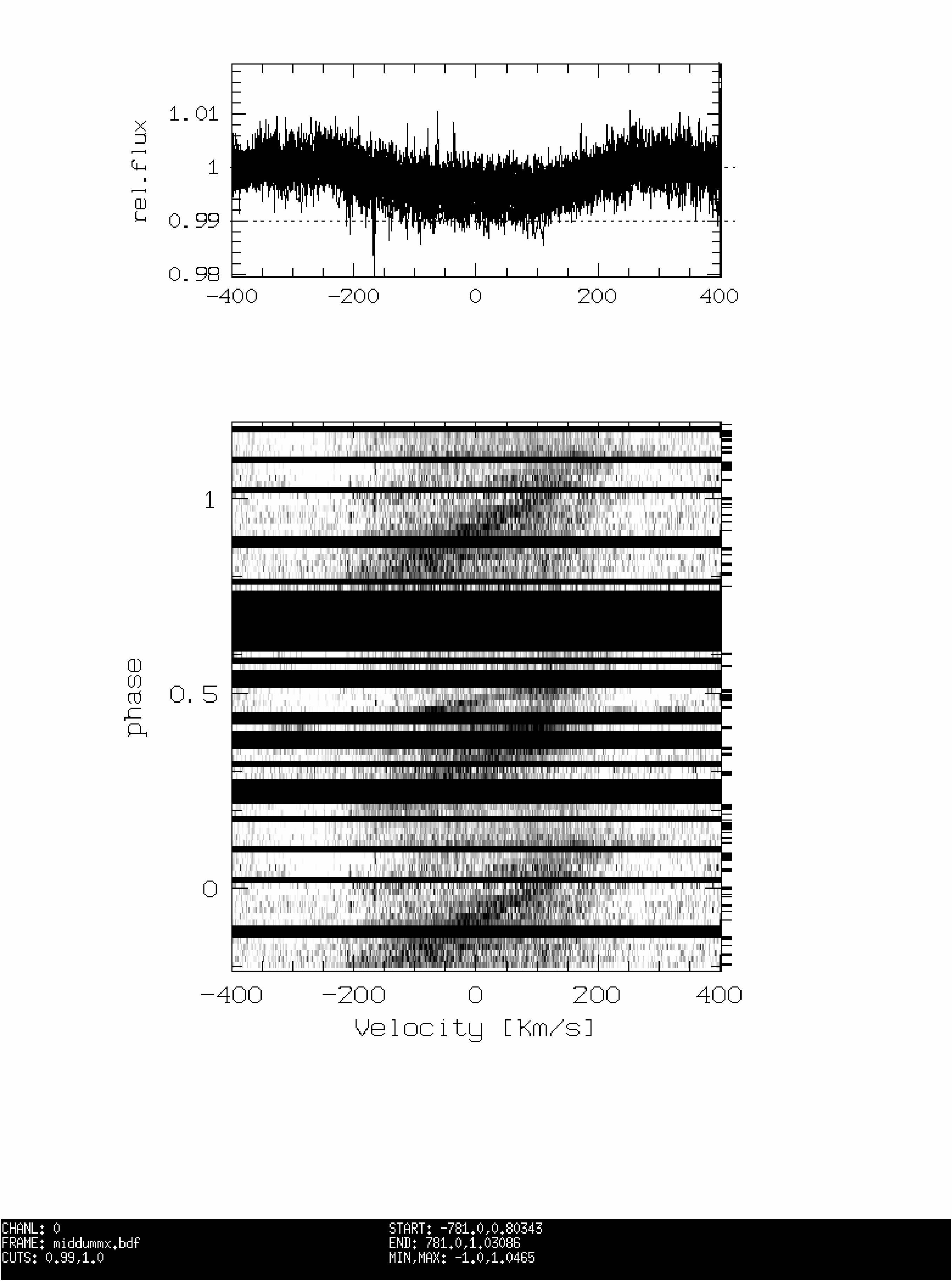}%
\includegraphics[viewport=100 188 608 1010,angle=0,width=0.25\textwidth,clip=]{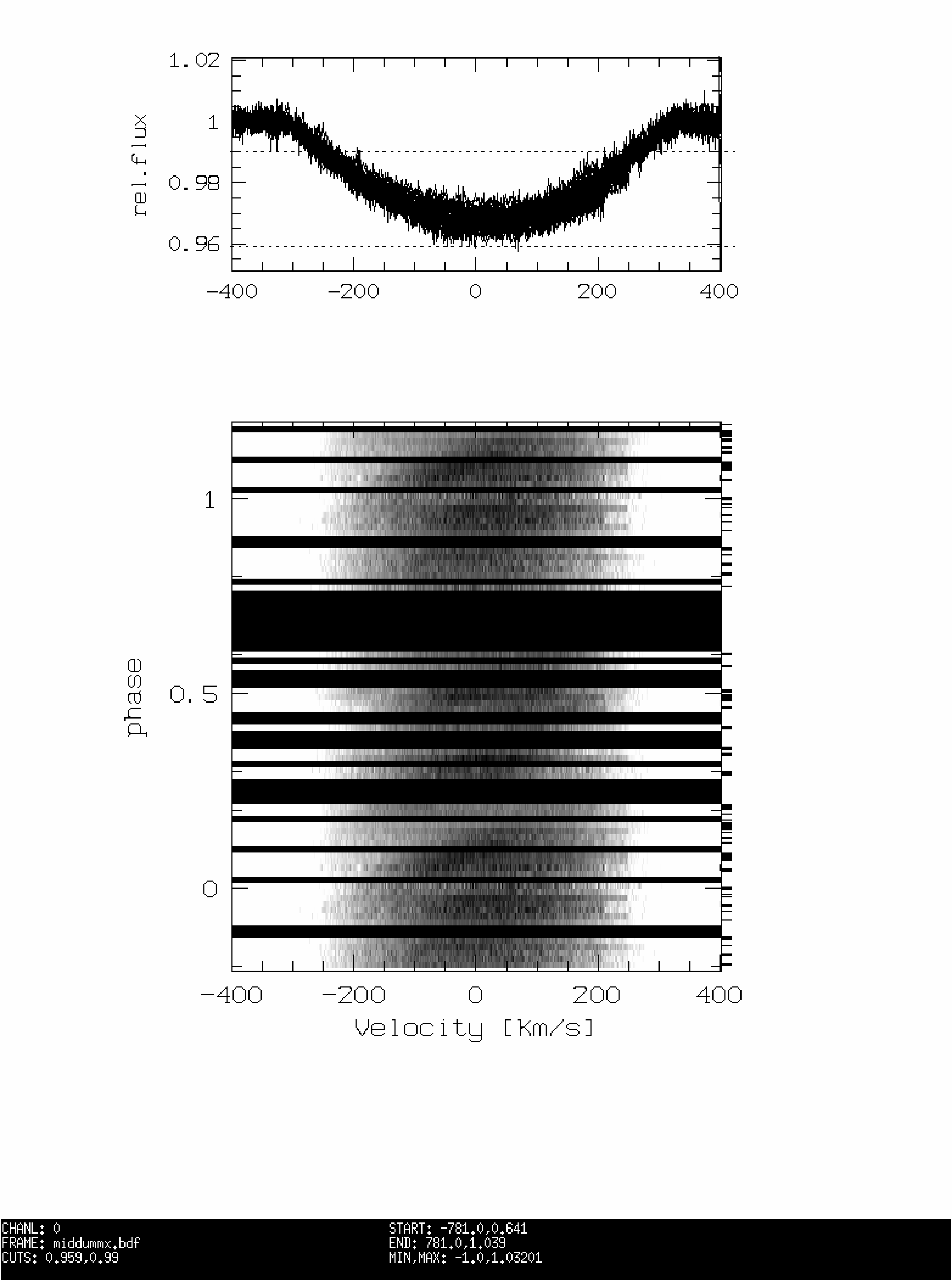}%
\includegraphics[viewport=100 188 608 1010,angle=0,width=0.25\textwidth,clip=]{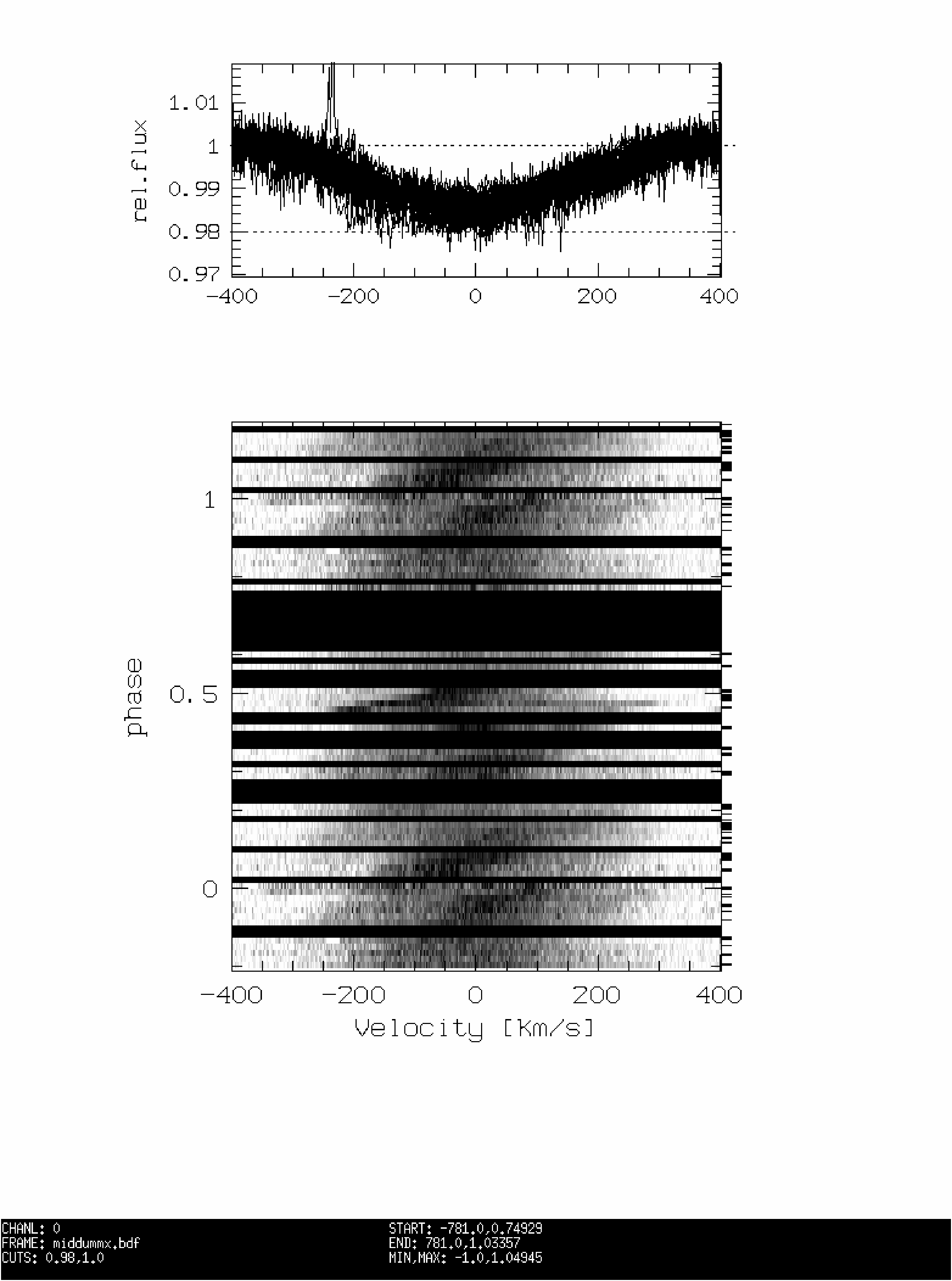}%
\includegraphics[viewport=100 188 608 1010,angle=0,width=0.25\textwidth,clip=]{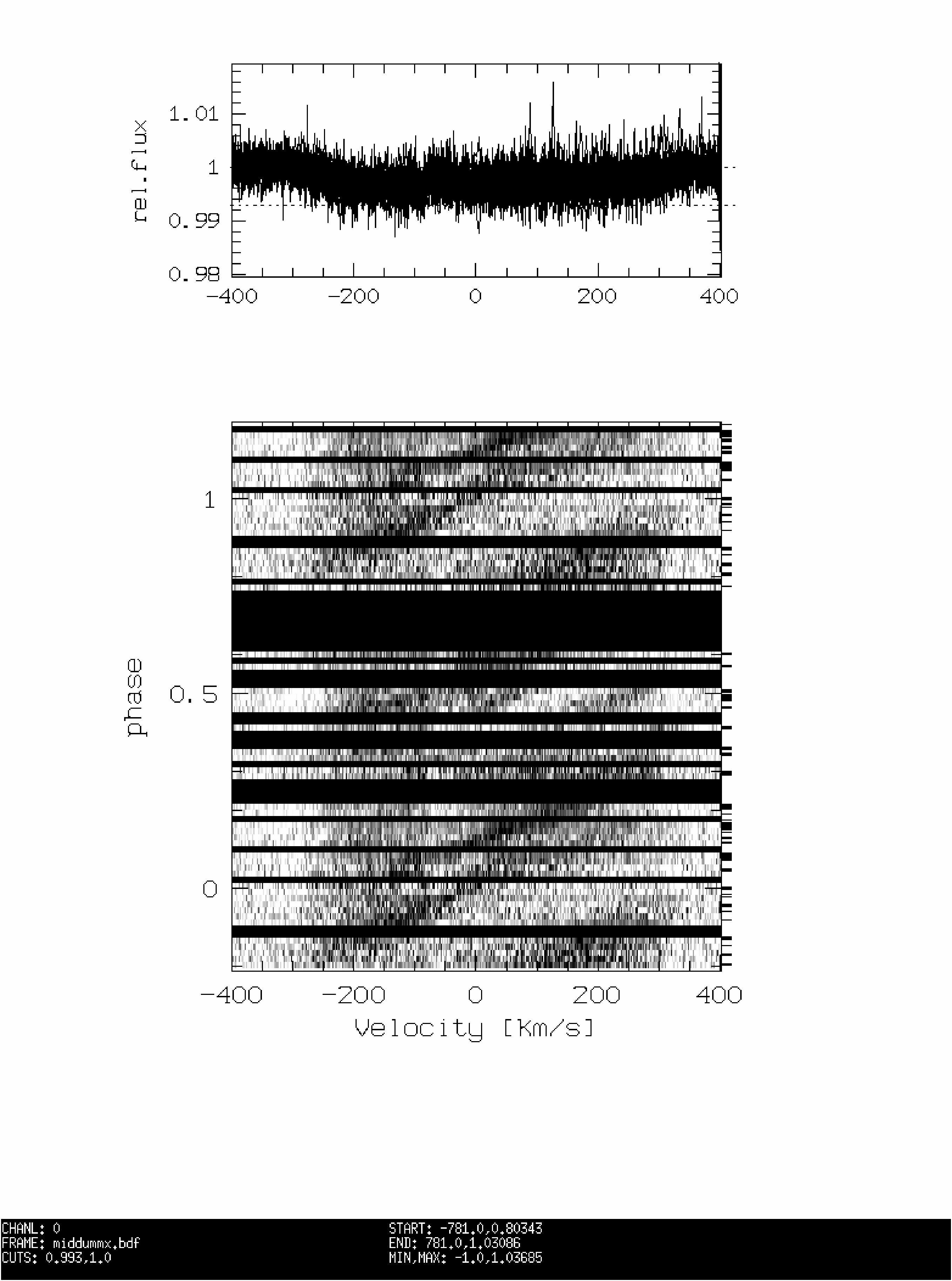}%

\end{center}
\caption[]{
\centering
Phased variations of photospheric Helium and metal lines
constructed with 64 bins.
}
\label{fig_lpv}
\end{figure}

The meaning of the {\bf effective temperature} $T_{\rm eff}$ in a rapidly
rotating star is not straightforward, so we note that $T_{\rm eff}$ here is
the uniform black-body temperature, that a star of the same surface area would
need to have the same full solid angle luminosity as the actual gravity
darkened star.  The $T_{\rm eff}$ of HR\,7355 is constrained using de-reddened
flux-calibrated UV spectra (from IUE) and visual photometry.  These data are
fitted best with $T_{\rm eff}=17\,000$\,K (see Fig.~\ref{fig_fluxfit}).

This temperature seems incompatible with the spectral classification
of B2. However, due to chemical peculiarity the helium lines are much stronger
than they would be for solar abundances, biasing classifications towards the
spectral type with the strongest He{\sc i} lines, which is B2.

Since $P_{\rm rot}$ and $v \sin i$ are well constrained, the {\bf inclination}
effectively determines the stellar equatorial radius and the geometric
projection onto the line of sight, i.e.\ the area of the star seen. Together
with the above derived $T_{\rm eff}$ and the Hipparcos distance, the flux
levels of the SED for 17\,000\,K constrain the inclination, therefore.  For
$T_{\rm eff}=17\,000$ models with $i=60^\circ$ are just so compatible with the
flux curve for the closest possible distance; better fitting inclinations
result in less plausible parameters elsewhere.

Although this still does not constrain the {\bf stellar mass} in a
straightforward observational way, all the other parameters are sufficiently
known to leave only a narrow range of acceptable evolutionary track
masses. This is because we can, at least in first order, expect that the
luminosity of a star of a given mass does not change with its rotational
velocity.  The model has a luminosity of about $L_\star = 1000\,{\rm
  L_{\odot}}$, and such a luminosity is only compatible with a mass of about
$6\pm0.5$\,M$_{\odot}$, not with a higher one.

\section{Variability}

For the {\bf ephemeris} we adopt the epoch given by \cite[Rivinius et
  al.\ (2010)]{2010MNRAS.405L..46R}, $T_0 ({\rm MJD}) = 54\,940.33 $, which is
the mid-date of an occultation of the star by the magnetospheric material,
seen in the Balmer line spectroscopy, and as such very well defined. The best
period proved to be the value published by \cite[Oksala et
  al.\ (2010)]{2010MNRAS.405L..51O}, $P=0.5214404$\,d, who could rely on a
twice as long time-base. Figure~\ref{fig_phasingall} shows previously
published data together with equivalent width measured in the UVES data,
phased with the ephemeris given here.


The  H$\alpha$ equivalent  widths are  fully  dominated by  variations of  the
circumstellar  environment,  there is  no  indication  for a  photospherically
intrinsic  variation of  this  or  any other  hydrogen  line's strength.   The
equivalent width of helium lines like  He{\sc i}\,4713 and 4388 are clearly of
a double wave character. Although  the strongest EW points in both half-cycles
are on  a similar  level, the  weakest EW values  differ by  a factor  of two.

In terms of line profiles, the most obvious photospheric variation is that of
the He{\sc i} lines. The variations are very strong, in percentage of the line
strength much stronger than in other He-strong stars like $\sigma$\,Ori~E.
Figure~\ref{fig_phasingall} shows a factor of three to four between the weak
and strong states of He{\sc i} lines. Other than the figure suggests, however,
a detailed look at the spectra shows that the variation is not as plain as
consisting of two enhanced polar regions and a depleted belt only, in
particular not in the metal lines, which hardly show EW variations, but clear
line profile variability (see Fig.~\ref{fig_lpv}).

\section{Magnetic field and magnetosphere}

Constraints on both the inclination $i$ and the obliqueness $\beta$ can be
obtained from the observed properties of the magnetic field.  Applying
standard diagnostics of stellar magnetic fields gives $r = B_{\rm min} /
B_{\rm max} = \cos(\beta+i)/\cos(\beta-i)= -0.78 $.  This means that $\beta+i
\lesssim 140^\circ$, for which $i=\beta=70^\circ$.  Since
we have determined $i=60^\circ$ above, we conclude that both $i$ and $\beta$
are between 60 and 80$^\circ$, in such a way that $\beta+i$ is about 130 to
140$^\circ$. The field strength at the magnetic poles would then be 11\,kG.
As will be shown in the next paragraph things are probably more complicated
than in a pure dipole field, but the above values should at be at least a good
approximation.

The measured magnetic curve, when fitted with a sine, has an offset of
$+0.3$\,kG, magnetic nulls expected to occur at $\phi=0$ and $\phi=0.54$
(dotted lines in lowermost panel of Fig.~\ref{fig_phasingall}). Yet, the cloud
transits are observed to occur at $\phi=0.0$ and $\phi=0.46$ (dotted lines in
upper panels of Fig.~\ref{fig_phasingall}).  Similarly, the magnetic poles
pointing towards the observer would be expected for $\phi=0.28$ and $0.78$ in
case of a sinusoidal variation (dashed lines in lowermost panel of
Fig.~\ref{fig_phasingall}), but spectroscopically, i.e.\ by maximal
He-enhancement, are rather seen at $\phi=0.25$ and $0.68$ (dashed lines in
upper panels of Fig.~\ref{fig_phasingall}).  Both these offsets could be
explained with a misaligned magnetic dipole wrt.\ to the stellar center, which
would display in a non-sinusoidal variability curve of the magnetic
field. Unfortunately, there are not enough magnetic measurements to address
this point in detail.

\begin{figure}
\parbox{0.05\textwidth}{~}%
\parbox{0.28\textwidth}{\centerline{H$\alpha$}}%
\parbox{0.33\textwidth}{\centerline{H$\beta$}}%
\parbox{0.33\textwidth}{\centerline{Pa$_{14}$}}%

\includegraphics[viewport=96 156 600 828,angle=0,width=0.33\textwidth,clip=]{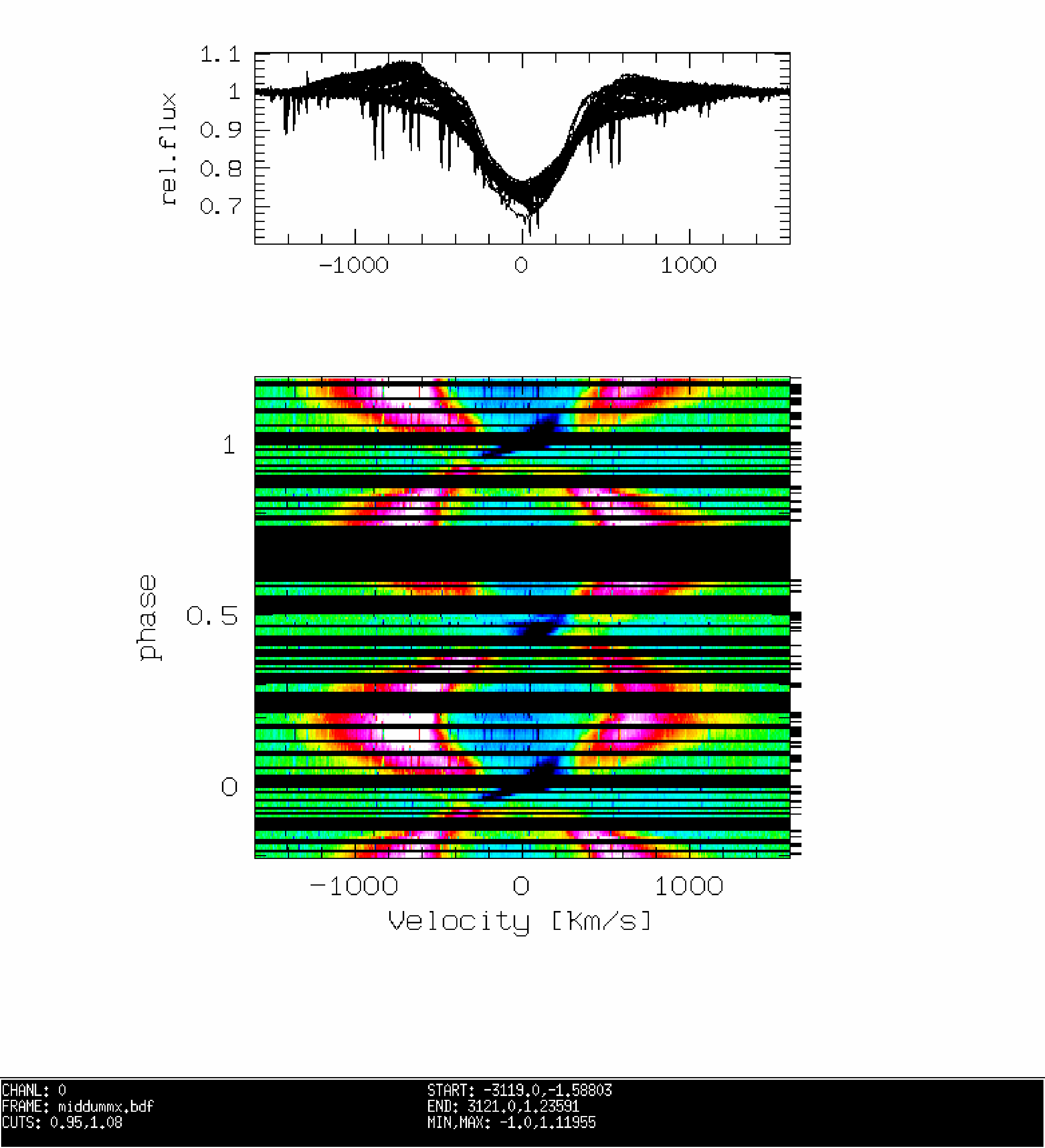}%
\includegraphics[viewport=96 156 600 828,angle=0,width=0.33\textwidth,clip=]{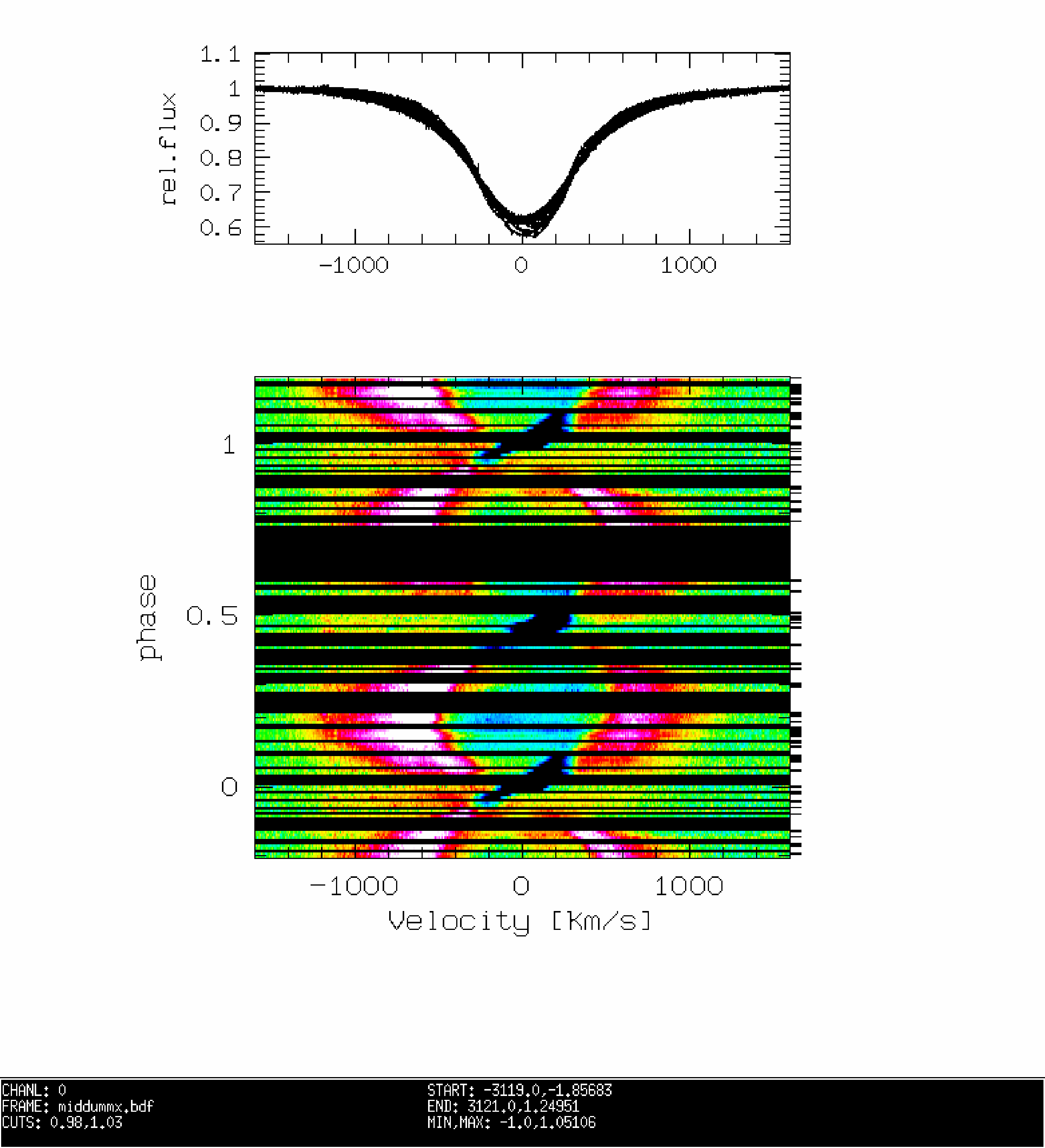}%
\includegraphics[viewport=96 156 600 828,angle=0,width=0.33\textwidth,clip=]{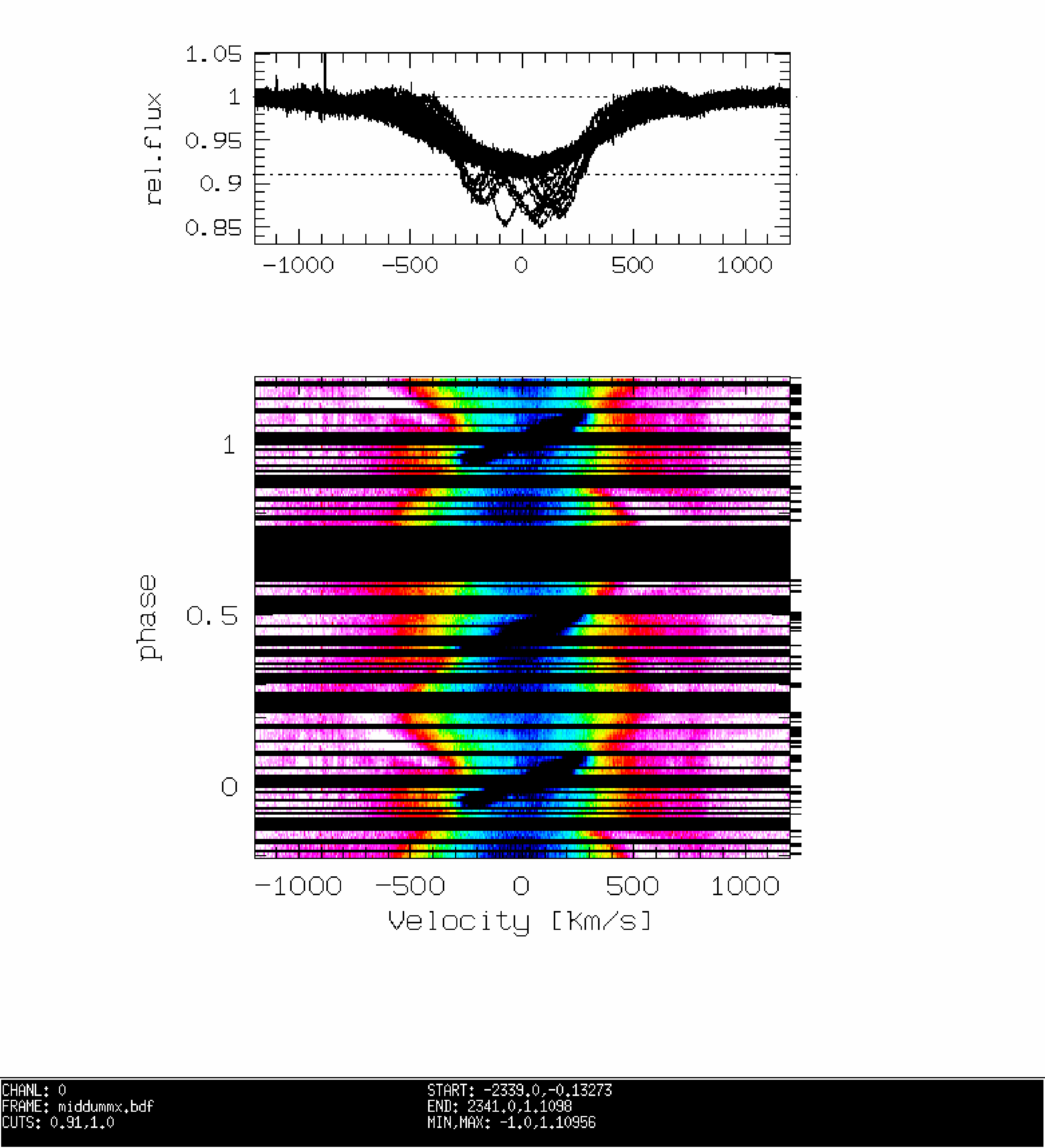}%

\caption[]{Colour-coded, phased observational variations of the circumstellar
  magnetosphere. H$\alpha$ and H$\beta$ are the residuals after subtracting
  the B3 model; the residual level in the line center is unity to better than
  1\,\%.  The respective original spectra are overplotted for H$\alpha$ and
  H$\beta$. Emission clearly seen in residuals, in fact up to
  H$\delta$. Paschen lines show emission clearly as well, shown here is
  Pa$_{14}$; note that these are not residuals and the different abscissa
  scale.  128 phase bins were used for all panels in order to sample the fast
  variations around the occultation phases in sufficient detail.}
\label{fig_dynamicalmagneto}
\end{figure}

The two occultations occur at phases $\phi=0.0$ and $\phi =0.46$
(Fig.~\ref{fig_dynamicalmagneto}).  The passing of the lobe is fast, about
$\Delta\phi=0.13$ from $-v \sin i$ to $+v \sin i$.  If the magnetospheric
lobes reached down to the star, this would rather be $\Delta\phi=0.5$. As the
crossing time decreases quadratically with distance, there is no absorbing
material directly above the star until about 2\,R$_{\star}$. A quarter of a
cycle later, the emitting material is seen next to the star, and since it is
in magnetically bound corotation there is a linear relation between velocity
and distance from the stellar surface.  The emission has an inner edge at
about $600$\,km\,s$^{-1} \approx 2 v \sin i$, which again points to an empty
region inside 2\,R$_{\star}$. The outer edge of the emission is at a velocity
of about $4\times v \sin i$ in H$\alpha$ and $3\times v \sin i$ in Pa$_{14}$,
which gives outer geometrical limits for the lobe emission of about
4\,R$_{\star}$ and 3\,R$_{\star}$, respectively.

The theoretical line profiles from the B3 modeling are good approximations of
the photospheric profile, as seen in Fig.~\ref{fig_fluxfit}. The residuals
(Fig.~\ref{fig_dynamicalmagneto}) represent the clean circumstellar emission
and it is possible to measure Balmer decrements when the emitting material is
next to the star. While the values for $D_{54}$ would be in agreement with
logarithmic particle densities between 11.7 and 12.5 per cm$^3$, the densities
from $D_{34}$ are somewhat higher, at 12.2 to 12.8 per cm$^3$. In any case,
these values are close to the optically thick limit, above which the
decrements become independent of density.

\section{Conclusions}

Apart from its high $v\sin i$, HR\,7355 is a rather typical member of the
class of He-strong stars. Due to its relative proximity and brightness it is
an ideal target to study the interplay of rapid rotation effects, like gravity
darkening and meridional circulation, with magnetic effects, like chemical
peculiarities and inhibition of the meridional circulation and other non-rigid
motions. Indeed it seems very unlikely that a relatively strong and ordered
field as seen in HR\,7355 could give rise to so many distinct zones of Helium
and metal abundances.

The presented data, more than 100 high-quality spectra filling the phase
diagram, provide an excellent base to apply techniques like Doppler imaging,
rigidly rotating magnetosphere models, and Monte-Carlo radiative transfer
models.

%
%
%

\begin{discussion}

\discuss{Groh}{Could you comment on the expected angular extension of the
  H$\alpha$ and Br$\gamma$ line forming regions and whether those could be
  resolved by interferometric observations?}  
\discuss{Rivinius}{Due to the rigidly rotating magnetosphere, $R_{\rm
    max}=v_{\rm max}/(v \sin i)$, so for both regions we expect about 3 to 4
  stellar radii. For HR\,7355 this is well sub-milliarcsecond, but for
  HR\,5907 (see Grunhut et al., this volume) at least the spectrally dispersed
  phase signature is probably in reach of AMBER/VLTI and the CHARA combiners.}

\discuss{Wade}{Due to the peculiarities hydrogen may be non-uniformly
  distributed in the atmosphere. Could the apparent departures from dipolar
  field topology be due to an effect like this?}

\discuss{Rivinius}{I would be surprised, since a) already the timing of the
  magnetosphere occultations, i.e.\ without looking at the magnetic field at
  all (except polarity) requires a non-sinusoidal (i.e.\ non-dipole) field
  signature and b) also with the ESPaDOnS data, derived from metal and
  He-lines, a similar argument can be made.}

\end{discussion}

\end{document}